\RequirePackage{fix-cm}
\documentclass[natbib,smallextended]{svjour3}       
\smartqed  
\usepackage{graphicx}
\usepackage{bbold}
\usepackage{graphicx}
\usepackage{hyperref}
\usepackage{amsmath}
\hypersetup{
    colorlinks,%
    citecolor=blue,%
    linkcolor=blue,%
    urlcolor=blue
}

\usepackage{ulem}

 \journalname{my journal}
\newcommand{\disrupchap}{Disruption Chapter}
\newcommand{\flowchap}{Formation of the Accretion Flow Chapter} 

\newcommand{\fmodchap}{Future Modeling Chapter}

\begin{document}

\title{Simulations of Tidal Disruption Events}



\author{Giuseppe Lodato       \and Roseanne M. Cheng   \and
        Cl\'ement Bonnerot \and Jane Lixin Dai 
}


\institute{G. Lodato \at
              Universit\`a degli Studi di Milano \\
              Dpartimento di Fisica\\
              Via Celoria 16, Milano (Italy)\\
              \email{giuseppe.lodato@unimi.it}           
            \and
            R.~M.~Cheng \at
              Theoretical Division (T-1, T-3)\\
              Los Alamos National Laboratory\\
              P.O. Box 1663\\
              Los Alamos, NM 87545 (USA)\\
              \email{rmcheng@lanl.gov}
           \and
           C. Bonnerot \at
              TAPIR\\
              Mailcode 350-17\\
              California Institute of Technology\\
              Pasadena, CA 91125 (USA)\\
              \email{bonnerot@tapir.caltech.edu}
            \and
              J.~L.~Dai \at
              Department of Physics\\ 
              The University of Hong Kong\\ Pok-fulam Road, Hong Kong\\  \email{lixindai@hku.hk}
}

\date{Received: date / Accepted: date}

\maketitle

\begin{abstract}
Numerical simulations have historically played a major role in understanding the hydrodynamics of the tidal disruption process. Given the complexity of the geometry of the system, the challenges posed by the problem have indeed stimulated much work on the numerical side. Smoothed Particles Hydrodynamics methods, for example, have seen their very first applications in the context of tidal disruption and still play a major role to this day. Likewise, initial attempts at simulating the evolution of the disrupted star with the so-called affine method have been historically very useful. In this Chapter, we provide an overview of the numerical techniques used in the field and of their limitations, and summarize the work that has been done to simulate numerically the tidal disruption process.
\keywords{black hole physics --- galaxies: nuclei --- hydrodynamics --- methods: numerical}
\end{abstract}

\section{Introduction}
\label{intro}

The dynamics of tidal disruption events is relatively complex. It involves treating the hydrodynamics of a self-gravitating fluid (the star), subject to a general relativistic force provided by the black hole. A full treatment of the problem would thus involve a general relativistic (magneto)-hydrodynamics code including heating and radiation.  In some respects, this is what is also required when describing accretion flows around black holes in other systems, such as Active Galactic Nuclei, X-ray Binaries, and Ultra Luminous X-ray sources.  Here the situation is made more complex because of the rapid variability of the system and the process of disruption, which is composed of several stages that are best described separately.  

In a nutshell, we can describe a TDE as being composed of three separate phases: (a) the disruption phase, (b) the evolution of the disrupted stream leading to disc formation, and (c) the accretion phase. In the disruption phase, the relevant physics is essentially the stellar self-gravity, fluid dynamics, and the tidal field of the black hole. For highly penetrating events, the tidal compression can lead to nuclear detonation (at least in the case of the disruption of a white dwarf), introducing complex thermodynamics in the problem. This phase is the simplest one, providing the only firm analytical prediction for TDE, the $t^{-5/3}$ fallback rate predicted by \citet{rees88} and \citet{Phinney89}. However, when treating it numerically, the problem arises that, in the inertial frame of the black hole, the action is spread over a very large volume, which is mostly composed of ``empty space'', and the tidal debris only occupies a relatively small portion of it. This is best treated either with a grid following the star for an Eulerian hydrodynamics code, or by using a Lagrangian hydrodynamics code. The second phase has been long thought to be relatively straightforward, with the debris essentially orbiting in Keplerian orbits until they fall back to pericenter. Recent investigations have shown complications as the flow can become unstable to self-gravity effects \citep{coughlin2015}. Subsequently, it is thought that the debris falls back and forms an accretion disc, although the process of disc formation is much less well understood (see the dedicated Chapter by Bonnerot et al. within this book). The process of disc formation can be strongly affected by both relativistic (such as apsidal and Lense-Thirring precession) \citep{ Dai15, Guillochon15, Hayasaki16, Bonnerot17} and magnetic effects \citep{Bonnerot17b}.  Finally, as mentioned above, the accretion phase is in some respects similar to the case of accretion tori in other black hole systems except that the conditions are extreme in terms of fallback rate.  Thus, we expect the formation of winds, outflows, and jets because this rate is strongly super-Eddington.

While no attempt has been made to simulate the whole process in a single framework, there has been significant progress over the years to simulate the individual stages independently. In this Chapter, we focus on the techniques that have been used in this context and on their applications to the various phases of TDE. The Chapter is organised as follows. In Section \ref{sec:techniques}, we describe the various kinds of codes that have been used for TDE: from Smoothed Particles Hydrodynamics, to grid-based methods, to the affine model that was specifically developed in the '80s to treat this problem. In Section \ref{sec:rel}, we describe the efforts that have been done to treat the important relativistic effects at play in the process. In Section \ref{sec:rad}, we describe the attempts at including more complex thermal and radiation physics. In Section \ref{sec:stages}, we review the numerical work that has been done to simulate the various phases of a TDE: disruption, stream evolution and accretion. Finally, in Section \ref{sec:conclusions}, we conclude and provide an outlook for the future simulations. 

\section{Numerical techniques used in TDE simulations}
\label{sec:techniques}

\subsection{Smoothed Particle Hydrodynamics}
\label{sec:sph}

Smoothed Particles Hydrodynamics (SPH, \citealt{lucy77,gingold77}) is a mesh-less Lagrangian algorithm that has been widely used, especially in cosmological simulations and hydrodynamic simulation of accretion discs and star formation. Its application to simulations of TDE is convenient especially in view of its Lagrangian approach, given the large dynamical range of the disruption phase of a TDE and since most of the computational domain is actually ``empty'', with the stellar debris occupying only a limited volume.  

It is interesting to note that some of the earliest application of SPH in astrophysics were indeed in the field of TDE, indicating that it was immediately apparent that a Lagrangian method such as SPH was particularly suited for this kind of problem. Looking back at the historical first papers, \citet{Nolthenius82} used $\sim 40$ SPH particles to simulate the disruption of a $1M_{\odot}$ star by a $10^4M_{\odot}$ black hole. \citet{Bicknell83} used 500 SPH particles to simulate a highly penetrating encounter of a $1M_{\odot}$ star by a $10^5M_{\odot}$ black hole, in order to test the possibility of tidal detonation of the star--a fashionable topic at the time. It is quite impressive to see how these early results were obtained with such limited numerical resolution (by comparison, modern codes use $\sim 10^6$ particles, a factor of ten thousand more than these seminal studies).

In SPH, the fluid is discretized into finite mass elements (called ``particles''), whose properties are computed by suitable averages between neighbouring particles that lie within a ``smoothing region'' around it.  Within this region, particles that lie close to the edge have a progressively lower weight. For example, the density $\rho$ at the position of particle $a$ is given by:
\begin{equation}
    \rho(\boldsymbol{r}_a)=\sum_{b=1}^{N} m_{b} W\left(\boldsymbol{r}_a-\boldsymbol{r}_{b}, h\right),
\end{equation}
where $W$ is the so-called ``smoothing kernel'', which is typically a bell-shaped function with compact support and a size defined by the parameter $h$, called the ``smoothing length''. $N$ is the number of particles within the kernel, which are called the ``neighbours'' of particle $a$. Each particle \emph{must} have their own smoothing length, determined in such a way to keep the number of neighbours roughly constant and avoid unwanted sampling error fluctuations between particles. A typical choice is:
\begin{equation}
    h\left(\boldsymbol{r}_{a}\right)=\eta\left(\frac{m_{a}}{\rho_{a}}\right)^{1 / 3},
\end{equation}
where, with $\eta=1.2$ (for the standard choice of a cubic spline kernel, see \citealt{Price12}) one has roughly 60 neighbours per particle in three dimensions. Thus, SPH naturally has the advantage that resolution follows density, so that high density regions have relatively small smoothing length (and thus have a higher spatial resolution). SPH is naturally an ``adaptive'' method. 

The equation of motion of the particles is derived from the Euler-Lagrange equations obtained from a variational principle formulation of fluid dynamics \citep{Eckart60}. The fluid Lagrangian $L$ is:
\begin{equation}
    L=\int\left[\rho v^{2}-u(\rho, s)-\Phi\right] d V,
\end{equation}
where $\rho$ is the density, $v$ is the fluid velocity, $u$ is the internal energy, that depends only on density and on the specific entropy $s$, and $\Phi$ is the gravitational potential. This can be discretized immediately, giving:
\begin{equation}
    L=\sum_{b} m_{b}\left[\frac{1}{2} v_{b}^{2}-u_{b}\left(\rho_{b}, s_{b}\right)-\Phi_{b}\right],
\end{equation}
from which the Euler-Lagrange equation of motion follows, after a little algebra and by using the first principle of thermodynamics for a dissipationless fluid (see, for example, \citealt{Price12} for details):
\begin{equation}
   \frac{\mathrm{d} \mathbf{v}_{a}}{\mathrm{d} t}=-\sum_{b} m_{b}\left[\frac{P_{a}}{\rho_{a}^{2}} +\frac{P_{b}}{ \rho_{b}^{2}} \right]\frac{\partial W_{a b}}{\partial \mathbf{r}_{a}} - \frac{\partial \Phi}{\partial \mathbf{r}_{a}}.
\end{equation}
The above equation is valid only for the simple case in which the smoothing length is assumed to be constant (a more complete equation can be found in \citealt{Price12}), but it shows that the SPH equations indeed recover the fluid equations appropriately, whereby in the left-hand side we find the Lagrangian derivative of the velocity and on the right-hand side we find the pressure gradient and the gravitational force. 
Since they follow the Euler-Lagrange equations, the set of SPH particles represent a Hamiltonian system and as such conserve exactly and simultaneously linear and angular momentum, as well as energy (see \citealt{Price12} and \citealt{Springel10} for recent reviews).  

The derivation described above, however, is only valid for a dissipationless fluid and requires all fluid properties to be continuous and differentiable. Particular care should thus be given to the handling of shocks and discontinuities in SPH \citep{Price08}. In particular, to resolve a shock, SPH typically uses an artificial viscosity. It can be shown \citep{Lodato10}, in the absence of switches, the artificial viscosity scales as $\propto c_{\rm s}h$, where $c_{\rm s}$ is the gas sound speed. Thus, low density regions, which are characterized by a large smoothing length $h$, also suffer from, in principle, unwanted, large artificial viscosity. In most cases, however, such dissipation can be effectively limited by using suitable switches, such as the \citet{Morris97} switch, or the \citet{Cullen10} switch. This can be particularly important for simulations that try to follow the formation of a disc after a TDE because, at the beginning of the fallback phase, the gas density is bound to be low.  Great care should be taken in ensuring that the results are not affected by excessive numerical dissipation.

Another interesting aspect to consider is the inclusion of relativistic effects. SPH naturally lends itself to a fully general relativistic implementation (in a fixed metric), as shown by \citet{Monaghan01}.  Progress is in underway to implement this feature in available codes. The PHANTOM code \citep{Price17} already implements several GR effects as corrections to Newtonian dynamics and very recently a full general relativistic implementation has been implemented \citep{Liptai19}. These include Lense-Thirring precession around a spinning black hole (following the approach of \citealt{Nelson00a}) and apsidal precession around a Schwarzschild black hole \citep{tejeda13}, which has been already used in simulations of TDE \citep{Bonnerot16}. These aspects will be discussed more extensively in Section \ref{sec:rel} below.

\subsection{Grid based codes}
\label{sec:grid}
An alternative to SPH is solving the fluid equations on a mesh.  In such grid based codes, the physical domain is discretized into volumes.  In each volume, the conservative equations of mass, momentum, and energy are evolved in time as
\begin{align}\label{eqn:euler}
\frac{\partial \rho}{\partial t} + \vec\nabla \cdot (\rho \vec{v})
      \nonumber      & = 0,\\
      \frac{\partial \rho \vec{v}}{\partial t} + \vec\nabla \cdot ( \rho \vec{v}\vec{v} ) + \vec\nabla P 
      \nonumber      & = 0,\\
      \frac{\partial E}{\partial t} + \vec\nabla \cdot [ (E + P) \vec{v}]
            & = 0,
   \end{align}
for density $\rho$, velocity $\vec{v}$, gas pressure $P$, and total energy density $E = u + \frac{1}{2} \rho |\vec{v}|^2$, where the internal energy density $u$ is determined by the choice in equation of state 
\citep{Land1959}.  For an ideal gas of adiabatic index $\gamma$, this is $u = P/(\gamma-1)$.
Extending to ideal magnetohydrodynamics, for magnetic field $\mathbf{B}$, the evolution equations are also in conservative form,
\begin{align}\label{eqn:euler_MHD}
\frac{\partial \rho}{\partial t} + \vec\nabla \cdot (\rho \vec{v})
      \nonumber      & = 0,\\
      \frac{\partial \rho \vec{v}}{\partial t} + \vec\nabla \cdot ( \rho \vec{v}\vec{v} - \vec{B}\vec{B} + \vec{P}^* ) 
      \nonumber      & = 0,\\
      \frac{\partial E}{\partial t} + \vec\nabla \cdot [ \ (E + P^*) \vec{v} - \vec{B} ( \vec{B} \cdot \vec{v})]
\nonumber            & = 0,\\
            \frac{\partial \vec{B}}{\partial t} - \vec\nabla \times (\vec{v} \times \vec{B})
            & = 0,
   \end{align}
where $\vec{P}^*$ is a diagonal tensor with components $P* = P + B^2/2$ and total energy density $E = u + \frac{1}{2} \rho |\vec{v}|^2 + \frac{1}{2}\vec{B}\cdot\vec{B}$ \citep{Gardiner2008,Stone2008}.  Note that special numerical treatment is necessary in order to enforce the solenoidal constraint $\vec\nabla\cdot\vec{B} = 0$ on the computational domain \citep{Evans1988}.  For a fixed region, decreasing the volume size increases the resolution of the simulation.  This method is useful in providing a high-resolution shock capturing technique to model, with high-accuracy, the dynamical response of the star to the black hole's tidal field as well as the formation and evolution of an accretion disc from the stellar debris.  It is inefficient when there is a significant amount of ``empty-space" in the simulation.

Unfortunately, this is the case for a significant phase of the TDE, namely when the star initially passes by the black hole.  Specifically, if one used a computational domain as large as the size of the expected stream, a majority of the region would be modeled as vacuum until a significant time after the disruption, when the thin stream of debris returned to the black hole, forming a disc.  By focusing the simulation on the star, adopting a reference frame with respect to its center, modeling ``empty-space" is avoided. Also, since the computational error scales linearly with the fluid velocity with respect to the grid, such a reference frame also helps in keeping the accuracy of the simulation high.  This type of calculation is useful for studying the dynamics of tidal compression and calculating quantities such as the return rate of debris to the black hole. Inefficiencies aside, grid based techniques are useful for modeling the disc formation and accretion not only because of the accuracy in capturing shocks, but also because implementing additional physics such as general relativity, magnetic fields, and radiation is straightforward.  Furthermore, many of these numerical tools are available to the astrophysics community due to the effort in solving related complex multi-physics problems in stellar structure and accretion discs.

To solve the basic equations of hydrodynamics, finite volume methods are often implemented because they are conservative by construction \citep{Toro1999}.  The numerical scheme is based on the Godunov update from time level $t^n$ to $t^{n+1}$, where the conservative equations of hydrodynamics Eq.~\ref{eqn:euler} and magnetohydrodynamics Eq.~\ref{eqn:euler_MHD} are discretized as
\begin{equation}\label{eqn:Godunov}
\mathbf{U}^{n+1}_{i} = \mathbf{U}^n_{i} + \frac{\Delta t}{\Delta x} \left ( \mathbf{F}^{n}_{i-1/2} - \mathbf{F}^{n}_{i+1/2} \right ),
\end{equation}
in one dimension over a zone of length $\Delta x$ within a time step $\Delta t$.  Mass, momentum, and energy $\mathbf{U} = \{\rho, \rho v^x, E\}$ are volume-averaged and defined at zone centers $i$ and fluxes $\mathbf{F} = \{ \rho v^x, \rho v^x v^x + P, (E+P)v^x\}$ are area-averaged and defined at zone faces $i\pm 1/2$.  At $t^n$, $\mathbf{U}$ is given.  The fluxes are time-averaged and depend on the solution to the Riemann problem or the jump discontinuity conditions of the conservative hydrodynamics equations at the zone faces.  High spatial accuracy of these schemes depends on the order chosen in solving the Riemann problem given the data defined at the zone-center and approximated at the zone face. The choice of numerical integration determines the temporal accuracy, where more sophisticated updates at intermediate time-levels may be implemented instead of Eq.~\ref{eqn:Godunov}, which is a first-order scheme.

For local simulations, the gravitational effects of the black hole are treated with respect to the local reference frame of the star.  Numerically, calculations of the disruption phase in this frame have greater accuracy than those with computational domains in the black hole frame.  This is because, when the star is tidally deformed, changes in the fluid velocity are much smaller than the orbital velocity of all of the debris with respect to the black hole.  These small changes are easily tracked in a local frame while they can be lost as numerical errors reduce the precision of calculations in the black hole frame.  Local, Newtonian treatments model the disruption of the star with Newtonian stellar and black hole gravity and ideal gas hydrodynamics \citep{Khokhlov93a,Khokhlov93,Brassart2008,Guillochon13}.  Post-disruption, the debris expands significantly and, as the density decreases, the hydrodynamical behavior becomes unimportant.  In this limit, the debris can be treated as ballistic particles.  If the local simulation follows the debris to this point, then an estimate of the return rate of debris to the black hole can be derived from the numerical results \citep{Guillochon13}.  While the star is typically modeled as a polytrope given by a single adiabatic index, the use of stellar profiles derived from the open source code MESA, Modules for Experiments
in Stellar Astrophysics, enables a study of the effects of tidal disruptions on stellar cores and tidal stripping of stellar envelopes \citep{MacLeod2012,MacLeod2013, Law-Smith2017,Anninos2018,LawSmith19,Goicovic19}. It is interesting to note that for some partial disruptions, the remnant core receives a ``kick'' into an unbound orbit \citep{Manukian2013}.
Furthermore, with a magnetized star, one can see the magnetic field amplification and structure in the debris field due to the tidal distortion \citep{Guillochon2017}.

In global simulations with respect to the black hole reference frame, the returning debris stream wraps around the black hole, forming a disc.  If the debris reaches within a radius significantly close to the black hole, then the fluid velocity is a significant fraction of the speed of light.  In this limit, it is useful to implement a fully general relativistic hydrodynamics (GRHD) code in order to accurately capture important general relativistic effects \citep[e.g.,][]{Haas2012}.  Furthermore, it is necessary to follow the interaction of gas with electromagnetic fields in the strong gravitational field around the black hole.  Plasma often possesses relatively strong magnetic fields.  Modeling their interaction is important for accretion and relativistic jet production.  Simulations of magnetic fields in the strong gravity regime are performed with general relativistic magnetohydrodynamic (GRMHD) codes \citep[e.g.,][]{Gammie2003}.  These codes are usually simplified with the use of an ideal gas equation of state, ideal MHD, and a static metric around the black hole.  For the latter, this works well for TDEs because the accretion of the debris from a star only trivially changes the mass and angular momentum of a supermassive black hole.  The equations evolved are similar to Eq. \ref{eqn:euler_MHD} but with relativistic stress-energy tensors, which we will explain in detail in Section \ref{sec:rel}. Due to the set up, GRMHD codes are more efficient when following the evolution of the gas after it has already formed a somewhat axisymmetric structure around the black hole \citep[e.g.,][]{Sadowski16}. 
Under certain circumstances the interaction between strong radiation and gas also needs to be well captured, such as when studying the super-Eddington accretion phase of TDE discs \citep[e.g.,][]{Dai18}. Such simulations can be performed using general relativistic radiation magnetohydrodynamic (GRRMHD) codes \citep[e.g.,][]{Sadowski14, McKinney14}, which evolve radiation in parallel with gas while including basic scattering and absorption physics.

We briefly discuss the current state-of-the art.  For local simulations, finite volume method codes such as \textsc{Flash} \citep{Fryxell2000} and \textsc{VH-1} \citep{Hawley2012} follow the star through its disruption phase in the Newtonian \citep{Guillochon2009} and post-Newtonian \citep{Cheng2013} regimes.  Before the disruption point, the star is a compact ball, but once it is disrupted, the gas expands, following trajectories around the black hole that further spread out the debris. To follow the debris long enough until the hydrodynamic effects are insignificant (at least for the unbound portion of the debris), all of the gas must remain inside the grid.  For a fixed Cartesian grid, one needs to start with a very large and computationally inefficient domain which models the ``empty-space" around the compact ball for a significant, and costly, amount of time.  In order to ameliorate this problem, \citet{Guillochon2009,Guillochon13} use an adaptive mesh refinement (AMR) technique \citep{Berger1989} which provides overlaying levels of increasing grid resolutions. Much of the debris is captured with a reduction of cost by placing high resolution on a high density regions of the debris and low resolution on ``empty-space."  However, this technique introduces uncertainties in numerical errors through interpolation between fine and coarse levels of refinement and relies on arbitrary conditions to distinguish between low-density debris and numerical atmosphere.  Such AMR techniques on a Cartesian grid have also been implemented in GRHD simulations of the disruption and disc formation such as with \textsc{Maya} in \citep{Haas2012}, to reduce the problem of computational inefficiency.  It is worth noting that the problem of maintaining the solenoidal constraint for magnetohydrodynamics is non-trivial with AMR.  Furthermore, \citet{Anninos2018} use a combination of AMR and moving-mesh techniques to model the disruption phase with
\textsc{Cosmos++} \citep{Fragile2012,Fragile2014}. Alternatively, one can re-express the spatial coordinate system used in evolving the hydrodynamic equations in terms of modified coordinates which provide logarithmic spacing in radius (grid zones increase with increasing radius) along with an increase in resolution along the equatorial plane \citep{Gammie2003,Noble2009,Sadowski2015}.  This is done with \textsc{Harm3d} in \citet{Shiokawa2015}, with \textsc{Harmrad} in \citet{Dai18} and with \textsc{Koral} in \citet{Sadowski16}. 

In the following sections, we discuss grid-based general relativistic treatments to the black hole and hydrodynamics \citep{Haas2012,Cheng2013,East2013,Shiokawa2015}, GRMHD \citep{Sadowski16}, the use of nuclear networks \citep{Anninos2018}, and studies with radiation \citep{Jiang2016} as well as with both radiation and general relativity \citep{Dai18, Curd19}.  While these multi-physics simulations begin to address questions about the leading physical mechanisms, the problem of computational expense limits the use of grid-based codes.  

\subsection{Moving mesh codes}

Recently, moving mesh Lagrangian codes have been developed, where the volume is discretized, as in grid-based methods, but the cells are allowed to move, thus including Galilean invariance, as in SPH methods \citep{Springel10,Hopkins15}.  The volume discretization is performed via a Voronoi tessellation, where the various cells are allowed to move and adapt to the flow. While not perfectly Lagrangian in nature, the method maintains Galilean invariance, unlike other grid-based methods. At the same time, it allows one to solve the Riemann problem at the cell interfaces, providing better shock capturing. In some sense, such moving-mesh codes combine the best properties of both grid-based and SPH methods.

Such methods have been rarely used to simulate TDE. \citet{Mainetti17} compare moving-mesh codes to SPH and grid-based codes for the problem of partial tidal disruptions, finding a relatively good agreement between the three methods. \citet{Anninos2018} combine moving mesh techniques with general relativistic hydrodynamics and thermonuclear reactions.  More recently, \citet{Steinberg19} have used moving mesh codes for the study of deeply plunging events, while \citet{Goicovic19} have also considered the evolution of partially disrupted stars, focusing on the evolution of the stellar remnant after pericenter passage.  

Another interesting application of moving mesh codes has been done by \citet{Yanilewich19}, who simulate the evolution of the unbound debris and the associated radio emission.

\subsection{The affine model}
\label{sec:affine}
To model the tidal disruption process, it is necessary to solve complex systems of non-linear equations for the fluid and self-gravity of the star as well as the gravity due to the black hole.  Analytical models provide reasonable estimates under basic assumptions.  A useful approximation is the limit where the pericenter of the star's orbit is much larger than the radius of the star ($R_p \gg R_*$). 
If the orbital timescale is much shorter than the stellar dynamical timescale, then the tidal forces dominate over the internal forces of self-gravity and pressure gradients, compressing the star significantly.  An interesting observational consequence of severe tidal compression was suggested by \citet{Carter82}.  For stellar encounters with pericenters that are much smaller than the tidal radius of the black hole ($\beta = R_t/R_p \gg 1$), a short phase of high compression or ``pancaking" leads to orders of magnitude increases in density and temperature, enough to detonate a significant fraction of thermonuclear fuel.  For a main sequence star, such a temperature increase would set off helium combustion by a triple-$\alpha$ reaction.  More nuclear reactions are possible \citep{Carter83}.  While gas cools as it expands after the compression, $\beta$-decay of by-products can reheat and eject the gas.  Potentially, this reheating leads to yet another stage of energy release by self-sustaining hydrogen combustion through a hot C-N-O cycle or rapid proton capture process, assuming that helium combustion is efficient. Note, however, that such results of the affine model are not reproduced in 3D hydrodynamical simulations of the disruption of main sequence stars, and nuclear detonation has only been demonstrated for the case of white dwarf disruption.

Coupling a dynamical model with nuclear reactions would reveal promising observational signatures.  A useful, analytic description is given by the affine star model, where layers of constant density are followed as ellipsoids in a frame with respect to the center of mass of the star \citep{Carter82, Carter83,Carter85} and is described, as follows.  Assume that general relativistic effects are negligible such that $R_p \gg G M_{\rm bh} / c^2$.  The star is defined in terms of 12 variables: 9 parameters for a $3\times 3$ deformation matrix $q_{ij}$ and the position of the center of mass in the black hole's reference frame $\vec{X}_0$.  Consider the quantities in the local frame of the star.  Let $\hat{r}$ be the position vector of a fluid element in the initial configuration of the star, in spherical equilibrium.  Let $\vec{r}$ be the position vector of a fluid element during the evolution.  In the affine limit, the star is constrained by the linear relation,
\begin{equation}
r_i = q_{ij} \hat{r}_j
\end{equation}
The mechanics is governed by a Lagrangian function $\mathcal{L}$, with $\mathbb{q}=q_{ij}$,
\begin{equation}
\mathcal{L} = \mathcal{L}_I(\mathbb{q}) + \mathcal{L}_C(\vec{X}_0) + \mathcal{L}_E(\vec{X}).
\end{equation}
%
The external term is $\mathcal{L}_E = T_E - M_* \Phi_E$, which depends on the external coordinates of the stellar fluid elements with respect to the black hole frame $\vec{X}$.  Specifically, for mass of the star $M_*$, $T_E = \frac{1}{2} M_* \dot{X}_i \dot{X}_j$ is the kinetic energy and $\Phi_E(\vec{X}_0)$ is the Newtonian gravitational field of the black hole evaluated at the center of mass of the star.  The coupling term is $\mathcal{L}_C = \frac{1}{2} \mathcal{M}_* C_{ij} q_{ik} q_{jk}$, where $C_{ij}(\vec{X}_0) = \partial_i \partial_j \Phi_E$ is defined in terms of derivatives with respect to the black hole frame coordinates $\vec{X}$ and evaluated at the center of mass of the star $\vec{X}_0$.  This trace-free, symmetric tensor represents the tidal force due to the black hole.  The quantity $\mathcal{M}_* = \frac{1}{3} \int \hat{r}_i \hat{r}_j dM$ is the scalar quadrupole moment of the initial configuration for the star.  The internal term is $\mathcal{L}_I = T_I - \Omega - U$, where the kinetic energy $T_I = \frac{1}{2} \mathcal{M}_* \dot{q}_{ij} \dot{q}_{ij}$, the self-gravitational potential $\Omega = -\frac{1}{2} \int \int dM dM' [(r_i-r_i')(r_i-r_i')]^{-1/2}$, and the total internal compression energy $U$ are all in terms of the internal coordinates, only.  The Lagrangian equations of motion for the fluid elements are decomposed into evolution equations for the external momentum $P_i = M_* \dot{X}_i$,
\begin{equation}\label{eqn:affine_external_momentum_evolution}
\dot{P}_i = -M_* \frac{\partial \Phi_E}{\partial X_i} + \frac{1}{2} \mathcal{M}_* q_{lk} q_{jk} \frac{\partial C_{lk}}{\partial X_i},
\end{equation}
and for the internal momentum $p_{ij} = \mathcal{M}_* \dot{q}_{ij}$,
\begin{equation}
\dot{p}_{ij} = \mathcal{M}_* C_{ik} q_{kj} + \Pi q_{ji}^{-1} + \Omega_{ik} q_{jk}^{-1},
\end{equation}
where $\Pi = \int P dV$ is the volume integral over pressure.  These evolution equations depend on the assumptions for the trajectory of the star $\vec{X}_0$, the equation of state which determines the total internal compression energy $U$, and the total power input due to nuclear reactions $\dot{Q}$.  In this model, the timescales are too short for radiative transfer.  Assuming a polytropic gas with index $\gamma$, the evolution of $q_{ij}$ is specified by
\begin{equation}\label{eqn:affine_evol}
\ddot{q}_{ij} = C_{ik} q_{kj} + (\gamma - 1) \Psi ||q||^{1-\gamma} q_{ji}^{-1} - \frac{3}{2} (\gamma - 1) \Psi_* \int_0^\infty \frac{du}{\Delta} \left ( \mathbb{S} + u \mathbb{1} \right)^{-1}_{ki} q_{kj},
\end{equation}
where $S_{ij} = q_{ik} q_{jk}$, $\Delta = || \mathbb{S} = u \mathbb{1} ||^{1/2}$, $\Psi_* = \mathcal{M}^{-1} U_*$ is the initial (equilibrium) state of the quantity that couples the thermonuclear release  and evolves as 
\begin{equation}
\dot{\Psi} = \mathcal{M}_*^{-1} ||q||^{\gamma-1} \dot{Q}.
\end{equation}
Without nuclear reactions, $\dot{Q}=0$.

Consider the Carter \& Luminet model for an adiabatic gas and in the absence of nuclear reactions.  Given the analytic form of the evolution for the star in Eq.\ref{eqn:affine_evol}, it is straightfoward to apply approximations in regions where the star is close to pericenter.  For encounters where $\beta \gg 1$, when the star reaches a pericenter much closer to the black hole than the tidal radius, the tidal forces are much greater than those due to the pressure and self-gravity of the gas.  The star is approximately particles in free-fall in the black hole's gravitational field, evolved as
\begin{equation}\label{eqn:affine_evol_freefall}
\ddot{q}_{ij} \approx C_{ik} q_{kj}.
\end{equation}
Neglecting the minor corrections due to the star's initial rotation, the motion is roughly invariant under reflection in the orbital plane.  Specifically, the motion is decoupled between the action orthogonal and within the orbital plane.  For orthogonal motion, Eq.~\ref{eqn:affine_evol_freefall} is $\ddot{q} = - C q$, where $q=q_{33}$, and the star ``pancakes", compresses and flattens in the orbital plane.  Once sufficiently flattened, the pressure forces dominate over the tidal forces, reversing the compression.  At this point, under the adiabatic approximation, where there is no nuclear energy release, all of the internal energy of the gas is converted from the kinetic energy of the vertical, pancake motion.  From this, the maximum temperature $\Theta_{\rm max}$ increases from its initial temperature $\Theta_*$ by
$\Theta_{\rm max} \approx \beta^2 \Theta_*$.  Assuming a polytropic equation of state with adiabatic index $\gamma=5/3$, we have the following.  The maximum increase in the initial central density $\rho_*$ is given by $\rho_{\rm max} \approx \beta^3 \rho_*$.  The timescale for maximum compression $\tau_{\rm max}$, in terms of stellar dynamical timescale $\tau_*$, is $\tau_{\rm max} \approx \beta^{-4} \tau_*$.  The time delay $t_{\rm max}$ between the instant of passage through pericienter and the instant of maximum compression is $t_{\rm max} \approx \beta^{-2} \tau_*$.  Extending the affine model for nuclear reactions, \citet{Luminet89a,Luminet89b} explore the process of nucleosynthesis in deep, plunging disruptions and its impact on the dynamics of the stellar debris and the isotropic enrichment of the interstellar medium.  

The limitations of the affine model are due to its idealized hydrodynamical description of the star and linearization of the properties of the black hole's tidal field.  It is appropriate for the central bulk of the star, but not the outer layers (which are less dense than the center).  Because the hydrodynamic description of compression is simple, it does not account for non-linear terms in the tide, which will produce shock waves, reducing the increase in density and temperature.  Nonetheless, the affine model is qualitatively useful for nuclear reactions and, as we discuss in the next section, extensions to this model can deliver interesting input in the relativistic corrections to the tide and orbit.

\section{Treatment of relativistic effects in simulations}
\label{sec:rel}

Numerical techniques used in simulating the relativistic effects of the black hole are generally divided into two categories: local simulations centered on the star and global simulations with a computational domain that includes the black hole.  For local simulations, the general relativistic formalism was first presented in the context of the affine model \citep{Luminet83,Luminet85,Luminet86,Marck1996}.  This framework was later used in grid-based simulations of the disruption phase \citep{Diener1995,Frolov1994,Cheng2013,Cheng14}.   In particular, by assuming that the star's radius is much smaller than the separation from the black hole, the stellar trajectory is approximately a geodesic.  For a non-spinning black hole of mass $M$ described by the Schwarzschild spacetime with coordinates $X^\mu = \{t, r, \theta, \phi\}$, to first order, the equations of motion are
\begin{align}
\frac{dt}{d\tau} & 
= \nonumber \frac{\epsilon}{1-\frac{2M}{r}},\\
\frac{d\phi}{d\tau} &
= \nonumber \frac{l}{r^2},\\
\left ( \frac{dr}{d\tau} \right )^2 &
= \epsilon^2 - \left ( 1-\frac{2M}{r} \right ) \left ( 1+\frac{l^2}{r^2} \right ),
\end{align}
for specific orbital energy $\epsilon$ and angular momentum $l$.  Within this approximation, a local inertial frame, sometimes called the Fermi normal coordinate frame, is constructed with respect to the star's center of mass \citep{Marck1983}.  The tidal potential due to the black hole is represented as
\begin{equation}
\Phi_i^{\rm tidal} = \frac{1}{2} C_{ij} x^i x^j + \frac{1}{6} C_{ijk} x^i x^j x^k + \frac{1}{24} C_{ijkl} x^i x^j x^k x^l + \cdots,     
\end{equation}
where $C_{ij}$, $C_{ijk}$, and $C_{ijkl}$ are the quadrupole, octupole, and hexadecapole tidal tensors etc. and $x^i = \{\tau, x, y, z\}$ represents the coordinates in the star's local reference frame.  This is a convenient form because the black hole tide is simply applied to the star through external accelerations in a locally flat spacetime.  This enables the use of a Newtonian hydrodynamics code with Newtonian self-gravity and high-order relativistic corrections to the black hole tide.  The tidal tensors are obtained from the curvature or Riemann tensor $R_{\mu\nu\alpha\beta}$ in the black hole frame.  For a known spacetime such as Schwarzschild or Kerr, $R_{\mu\nu\alpha\beta}$ is given.  Using the Fermi normal frame vectors or tetrad $\lambda^\mu_{\ i}$, the Riemann tensor in the black hole frame is transformed into the local inertial frame, where the tidal tensors can be calculated and added to the tidal potential.  For example, for the quadrupole tensor $C_{ij}$ we have
\begin{equation}
C_{ij} \equiv R_{\tau i \tau j} = R_{\mu \nu\alpha\beta} \lambda^\mu_{\ \tau} \lambda^\nu_{\ i}\lambda^\alpha_{\ \tau} \lambda^\beta_{\ j},
\end{equation}
where $R_{ijkl}$ is the Riemann tensor transformed into the local star's frame.
Extensions to higher-order tides and spin effects are straightforward in this Fermi normal frame \citep{Ishii2005,Cheng2013}.

Relativistic corrections to the tides lead to more disruptive encounters than the Newtonian approximation for a given encounter strength \citep{Frolov1994,Diener1995}.  The relativistic effects are seen in the return rate through a slower rise and lower peak with respect to the Newtonian encounters \citep{Cheng14}.  Local simulations of the disruption phase are especially useful because at the end of the simulation, the orbital parameters of the debris are on-hand to either make predictions about the disc or use as initial conditions for a fully global simulation of the interacting debris streams and the formation of the disc \citep{Shiokawa2015}.

The full general relativistic form of the basic conservative hydrodynamics equations in a black hole spacetime defined by coordinate $X^\mu$ and black hole metric $g_{\mu\nu}$ is given by
\begin{align}\label{eqn:conservative_hydro_GR}
\partial_t (\sqrt{-g}\rho u^t) + \partial_i(\sqrt{-g}\rho u^i) 
\nonumber & = 0,\\
\partial_t (\sqrt{-g}T^t_{\ \mu}) + \partial_i(\sqrt{-g} T^i_{\ \mu}) 
& = \sqrt{-g}\Gamma^\sigma_{\ \rho\mu} T^\rho_{\ \sigma},
\end{align}
in terms of four-velocity $u^\mu$, stress-energy tensor $T^{\mu\nu}$, connection coefficient $\Gamma^\sigma_{\ \rho\mu}$, and determinant of the black hole metric $g$.  Extending to ideal GRMHD \citep{Gammie2003,White2016}, the stress-energy tensor is written in terms of the magnetic field components $B^i$, where
\begin{equation}\label{eqn:electromagnetic_stress-energy_tensor}
T^{\mu\nu} = (\rho h + b_\lambda b^\lambda ) u^\mu u^\nu + \left ( P + \frac{1}{2} b_\lambda b^\lambda \right ) g^{\mu\nu} - b^\mu b^\nu,
\end{equation}
for specific enthalpy $h$ and contravariant magnetic field components $b^0 = g_{i\mu} B^i u^\mu$ and $b^i 
 = ( B^i + b^0 u^i)/u^0$,
which define the dual of the electromagnetic field tensor as
\begin{equation}\label{eqn:electromagnetic_field_tensor}
    ^*F^{\mu\nu} = b^\mu u^\nu - b^\nu u^\mu.
\end{equation}
Then, in addition to Eq.~\ref{eqn:conservative_hydro_GR}, the conservative equations for evolving the hydrodynamic variables as well as the magnetic field are
\begin{equation}
    \partial_t (\sqrt{-g} B^i ) + \partial_j (\sqrt{-g}\  {^*F^{ij}}) = 0.
\end{equation}
Note that while the magnetic field $B^i$ is evolved, the electric field $E^i$ is inferred from $B^i$ and the fluid velocities $u^\mu$ through Eq.~\ref{eqn:electromagnetic_field_tensor} and the definition,
\begin{equation}
    {^*F^{ij}} = 
    \begin{pmatrix}
    0 & -B^1 & -B^2 & -B^3\\
    B^1 & 0 & E^3 & -E^2\\
    B^2 & -E^3 & 0 & E^1\\
    B^3 & E^2 & -E^1 & 0
\end{pmatrix}.
\end{equation}
To-date, there are few tidal disruption event simulations that solve the full general relativistic set of equations in the black hole frame.  This is because the computational expense for the hydrodynamics scales with black hole to star mass ratio.  Most grid-based black hole frame codes leverage the computational expense with a small black hole to star mass ratio.  In particular, white dwarf disruptions by intermediate mass black holes are chosen because of the computationally tractable length and timescale as well as the fact that the parameters are well within the relativistic regime without the use of a large penetration factor $\beta$.  To further decrease the computational expense, \citet{Shiokawa2015} consider the problem in a two-step process where post-disruption simulation data from \citet{Cheng2013} is used as initial data for a global GRHD simulation of the accretion flow of the bound stream.  This technique is also implemented in \citet{Sadowski16}, with initial data from \citet{tejeda13}, in a study with GRMHD.  Both studies use a fixed Schwarzschild spacetime.  The initial data represents the results of a hydrodynamic treatment of the star under the influence of its own gravity and the black hole.  

While the stellar self-gravity is important for the disruption process, it is much less so very close to the black hole.  A few grid-based studies which focus on deeply relativistic encounters neglect self-gravity in order to save the computational expense for different physical effects \citep{Haas2012,East2014,Anninos2018}.  \citet{Haas2012} use a fully general relativistic hydrodynamics code with an adaptive mesh refinement technique where both the star and the black hole share the computational domain.  Several refinement levels follow the star as it passes by a spinning black hole.  It is found that after disruption, the black hole spin disperses the debris scattering it far from the orbital plane.  \citet{East2014} models a more extreme mass ratio while making use of an error-correcting technique which reduces the expense of calculating the black hole and star's gravity.  Both codes used in \citet{Haas2012} and \citet{East2014} directly calculate the gravitational wave form.  The former provides the waveform for a white dwarf disrupted by an intermediate mass black hole.  The latter study focuses on generating the waveform for a main sequence star disrupted by a $10^6$ solar mass black hole.  \citet{Anninos2018} use a fully general relativistic hydrodynamics code in the black hole frame, but make use of a moving mesh technique as well as adaptive mesh refinement in order to focus the computational domain on the star only.  This decrease in computational expense allows more resources dedicated to resolving the star and coupling a nuclear network to the hydrodynamic evolution.  The results of this study, on thermonuclear reactions in tidal disruption events is discussed below.

The question of what gravitational wave form is generated from a tidal disruption event has been the focus of a wide variety of studies using the affine model \citep{casalvieri2006}, smoothed-particle hydrodynamics \citep{Rosswog09,Kobayashi2004}, and full general relativistic hydrodynamics \citep{Haas2012,East2014,Anninos2018}.  Tidal disruption events generate a burst-like gravitational wave with low frequencies of $0.1-0.7$ mHz and strain amplitudes of $h\sim 10^{-22}$ at a source distance of 10 Mpc. More recently, \citet{Toscani19} have also studied the gravitational wave emission generated by hydrodynamical instabilities in the accreting torus, after the disruption proper. The typical frequency in this case is similar to the case of disruption (in the mHz regime), the strain is lower $h\sim 10^{-24}$, but rather than an instantaneous burst, in this case the emission is expected to be produced over a few cycles, which may help detection.




Following the disruption of a star, the debris orbits around the black hole with apsidal and nodal precession due to general relativity \citep{Merritt13}, the rate of which dictates how fast the debris stream can form a disc from self-intersection. As discussed previously, simulating the motion and evolution of debris stream around the black hole with hydrodynamical codes is computationally inefficient even when such codes are Newtonian. However, under the assumption that hydrodynamical effects are unimportant for the debris orbital motion, the debris can also be approximated to follow geodesics around the black hole until debris stream self-intersection happens. In many cases, even semi-analytical calculations only considering the first-order precession rate are sufficient \citep{Dai15, Guillochon15, Bonnerot17}. One can also more accurately calculate the geodesics using a numerical approach. For example, \citet{Dai13} calculated the debris motion in full general relativity around a spinning black hole by solving the geodesic equations of motions from \citet{Fuest04}.

A computationally efficient way of treating general-relativistic effects in SPH simulations consists in using a gravitational potential that incorporates the most important ones. One possibility is to use pseudo-Newtonian potentials that reproduce the rate of relativistic apsidal precession around the black hole. For instance, the potential developed by \cite{Wegg12} reproduces this feature accurately for highly-eccentric orbits while that of \cite{tejeda13} provides an exact treatment of the gas trajectories around a non-rotating black hole. These approaches have been exploited to study partial stellar disruptions \citep{gafton2015} and disc formation \citep{Hayasaki13,Bonnerot16}. In the latter case, a correct evaluation of the apsidal angle is fundamental since it is at the origin of the self-crossing shock that initiates this process. Spin-related effects can be taken into account in a similar way by adding post-Newtonian acceleration terms to the Keplerian component as described by \citet{Blanchet06}. This method has been used to study the impact of Lense-Thirring precession on the stream-stream collision and ensuing circularization \citep{Hayasaki16}. In the above approaches, note that general relativity is only taken into account to specify the gravity of the black hole while the hydrodynamics still assumes a flat space-time.

A full general-relativistic treatment of the gas dynamics is also possible using SPH (see the recent implementation in the PHANTOM code, \citealt{Liptai19}), but it requires a rewrite of the equations of gas dynamics to account for the space-time curvature, as detailed in \citet{Rosswog09}. The earliest application of fully general relativistic SPH to tidal disruption events was provided by \citet{Laguna93}, using the code developed by \citet{Laguna93b} based on the approach of \citet{Kheyfets90}. This formulation, however, was not in conservative form (as the one described above in section \ref{sec:sph}) and had problems in resolving shocks efficiently. More modern SPH versions have been developed \citep{Rosswog09,Liptai19} that are fully conservative (thus exploiting the major advantage of an SPH formulation) and are able to handle shocks much more efficiently. The application of such methods to TDE has been limited, though. Note that this approach assumes that the metric is entirely specified by the black hole and not affected by the mass of the gas that evolves around it. Of course, this is a very good approximation for tidal disruptions where the compact object is always much more massive that the stellar debris. This technique has been used by \citet{Tejeda17} and more recently by \citet{gafton2019} and \citet{Liptai19b} to simulate the disruption of a star by a Schwarzschild and a rotating black hole. In these investigations, note that the gas self-gravity remains treated in a Newtonian way that, although generally being a excellent approximation, is strictly speaking not entirely self-consistent.

\section{Treatment of thermodynamics, radiation and nuclear reactions}
\label{sec:rad}

In the description of simulation methods above, the primary goal of previous work has been to study the dynamics of the stellar gas  through gravitational interactions. Generally, the stellar model is assumed to be a non-rotating polytrope (except in recent papers, such as \citealt{Sacchi19} and \citealt{Golightly19}), where the distinguishing characteristic between main sequence stars and white dwarfs, for example, is the difference in adiabatic index.  Radiation has not been included in most simulations 
although the importance of its role is still an open question.  While the interaction of the debris streams may reduce the high rate of the funneling of gas around the black hole, the efficiency of accretion is not well understood.  As new multi-wavelength observations are likely to deliver detailed light curves, it is increasingly important to have theoretical models which explain the spectral features.  This calls for new attention to the description of the stars themselves, their layers and nuclear constituents, as well as a direct treatment of radiative processes helpful in understanding the nature of multi-wavelength emission in TDE.  

The state-of-the art includes few studies of the thermodynamics beyond ideal gases.  Differences in the choice of adiabatic index of the gas lead to changes in the density structure of the star, which in turn affects the shape of the debris return rate \citep{Lodato09,Guillochon13}.  If the light curve of TDE traces the return rate, then these theoretical models could help determine the types of stars that are disrupted.  In detailed stellar structure studies, \citet{MacLeod2012,MacLeod2013} use initial configurations from a MESA stellar evolution code while evolving the star as a composite polytrope.  This enables a simulation code to model tidal stripping of material from envelopes and partially disrupt cores.  

Local radiation-hydrodynamics simulations have been carried out to evaluate the amount of radiative losses experienced by the heated gas after it has passed through the self-crossing shock at the origin of disc formation. In the early investigation of \citet{kim1999}, radiation transport is approximated by incorporating a volume cooling rate. The more recent work by \citet{Jiang2016} makes use of a more accurate treatment that consists in solving the frequency-independent radiation-hydrodynamics equations. 

Furthermore, for the evolution of the debris disc, choices in the adiabatic index of the gas lead to dramatically different configurations: thin, circularized (isothermal) and puffy, elliptical (adiabatic) \citep{Bonnerot16, Hayasaki16}. For GRMHD codes used to simulate the formation of accretion flow \citep{Shiokawa2015, Sadowski16}, the gas is usually assumed to have an adiabatic index of 4/3 (relativistic) or 5/3 (Newtonian). However, when simulating accretion discs that are optically thick where the dynamics of the gas and radiation are coupled, such as in the super-Eddington accretion regime, the sensitive interaction between gas and radiation needs to be captured. In order to address this, GR-radiation-MHD (GRRMHD) codes have been developed in recent years. The state-of-the art codes use the flux-limited diffusion approximation which only
allows isotropic emission relative to the fluid frame \citep[e.g.,][]{Ohsuga05} or the more accurate M1 closure \citep{Sadowski14, McKinney14, Dai18, Curd19} for the gas-radiation coupling. The stress-energy tensor evolved in the equations in these codes does not only contain the electromagnetic energy component and the matter component as in Eq. (\ref{eqn:electromagnetic_stress-energy_tensor}), but also a radiation stress-energy component. The dynamics of gas and radiation are evolved in parallel and coupled to each other using opacity terms addressing electron Thomson and Compton scattering as well as various absorption processes.

Motivated by \citet{Carter82} and \citet{Luminet89a,Luminet89b}, several numerical studies investigate the interesting observational consequences once the star reaches pericenter.  While most probe the effects at very close encounters with the black hole, a few  focus on weakly disruptive encounters in the relativistic regime \citep{Wilson2004,Dearborn2005}.  The main focus of studies is modeling the nuclear reactions triggered by severe compression in the orbital plane leading to an energy release comparable to Type Ia supernovae, particularly with white dwarf disruptions by intermediate mass black holes.  The threshold for nuclear ignition is set by the star's gravitational binding energy.  Given a rapid increase in temperature and energy above this threshold within a short timescale, the star is considered ignited.  Several numerical studies model the temperature increase due to the shock after severe compression \citep{Kobayashi2004, Brassart2008, Rosswog2009, Guillochon2009, Haas2012, TanikawaMar2018, TanikawaMay2018,Kawana2018, Anninos2018, Kagaya2019}.  \citet{Rosswog2009} model the compression and expansion during the disruption phase while tracking the energy and nucleosynthesis of the debris.  To track nucleosynthesis, they implement a \textsc{Helmholtz} equation of state with flexible specification of the chemical composition of the gas and couple an $\alpha$-network with seven nuclear species to the hydrodynamic evolution.  Several models meet the conditions for nuclear ignition with sizeable iron-group injection into the outflow.  However, \citet{TanikawaMay2018} finds a numerical limit to accurately resolve nuclear flows and notes that ignition is sensitive to resolution.  In a detailed follow-up investigation with general relativistic hydrodynamics, initial data from \textsc{Mesa}, a \textsc{Helmholtz} equation of state, and a large nuclear network, \citet{Anninos2018} show ignition with diverse thermonuclear environments and in particular, show calcium-group injection in the outflow in addition to the iron-group.

\section{Simulating the various stages of TDE}
\label{sec:stages}

\subsection{Disruption}
In this section, we provide an overview of the evolution of simulations of the disruption phase in TDE, emphasizing the numerical advances and innovations obtained in the various studies. Not all of the papers investigating the disruption phase are reported here (this is the subject of the \disrupchap{} in this book), but only those that have provided significant improvements from the point of view of the numerical techniques. 

After the seminal simulations by \citet{Nolthenius82} and \citet{Bicknell83}, the first simulations of the disruption process with reasonable resolution were performed by \citet{Evans89}. This paper is particularly important, as it sets the standard for the numerical simulations of the disruption process. They simulate the disruption of a $1M_{\odot}$ star by a $10^{6}M_{\odot}$ black hole, with a penetration factor $\beta=1$, on a parabolic orbit. They were the first investigators to derive the distribution of mechanical energies of the debris, apparently confirming spectacularly the prediction of a flat energy distribution by \citet{rees88} (although their plots were in log scale, which hides possible smaller changes in the distribution) and they obtain the predicted fallback rate, again confirming Rees' prediction of a $t^{-5/3}$ decline. From the numerical point of view, they use $\sim 10^4$ particles and perform a convergence test, showing that for this stage of the process, the simulations are converged in SPH. 

\citet{Laguna93} were the first to introduce relativistic effects in SPH simulations of the disruption process (see Section \ref{sec:rel}) and simulated strongly penetrating disruptions of a $1M_{\odot}$ star by a $10^6M_{\odot}$ Schwarzschild black hole. Their numerical resolution, however, was quite low, even for the standards of those days (they used $10^3$ particles). 

In a series of relatively poorly known, but pioneering papers, \citet{Khokhlov93a}, \citet{Khokhlov93} and \citet{Diener97} perform some of the first Eulerian simulations of TDE. They simulate the disruption of polytropic stars, with different polytropic indices, on a fixed and on a moving mesh. \citet{Khokhlov93a} simulate weak encounters that do not lead to disruption. \citet{Khokhlov93} simulate partial and total disruptions and estimate the critical penetration parameter to have disruption as a function of stellar structure. They find $\beta_{\rm crit}\sim 0.65-0.75$ for $\gamma=5/3$, and $\beta_{\rm crit}\sim 1.4-1.6$ for $\gamma=4/3$. Subsequent simulations with much higher resolution and AMR techniques by \citet{Guillochon13} later corrected these values to $\beta_{\rm crit}=0.9$ and $\beta_{\rm crit}=1.85$, for the two cases, respectively. \citet{Diener97} introduce full relativistic effects on a fixed metric in their Eulerian code, such that the debris evolves following Kerr geodesic. Interestingly, despite their pioneering quality, such papers did not achieve significant success in the community, probably because their main outcome was essentially an estimate of the amount of mass, energy and angular momentum transferred in the encounter. 

\citet{Ayal00} try to follow the evolution of the tidal debris both during the disruption phase and during the subsequent stream evolution. They use a post-Newtonian approach to include some relativistic effects into an SPH method. However, their simulations suffer significantly from poor resolution. While the number of particles used is similar, or slightly lower than those of their contemporaries (they employ $\sim 10^3$ particles), the resolution issue comes about because they want to study the post-disruption phase, where the debris significantly spread and reach very low densities, thus requiring a very large smoothing length in SPH. They attempt to resolve this issue by employing the so-called ``particle splitting'' technique that however introduces significant noise in the particle distribution and has been mostly abandoned in more modern SPH studies.

Two ``modern'' papers investigating the disruption phase are those of \citet{Lodato09} and of \citet{Guillochon13}. Both papers investigate the disruption of a solar mass star by a $10^6M_{\odot}$ black hole essentially using advanced versions of the codes employed in the past. \citet{Lodato09} use a modern SPH code with a full Lagrangan formulation, employing adequate viscosity switches and a variable smoothing length (that was not always used in the past) and employ a reasonably large number of particles ($10^5$), thus reaching adequate numerical resolution. \citet{Guillochon13} use the FLASH code, with AMR. Both use Newtonian gravity. Their main results can be summarized as follows. \citet{Lodato09} revisit the conclusions of \citet{Evans89} noting that the energy distribution of the debris is actually not exactly flat for the most bound debris, leading to deviations to the $t^{-5/3}$ law at early times in ways that depend on the internal structure of the star. Such deviations can be even predicted analytically, although the analytical calculations need some adjustments in order to reproduce the numerical results. \citet{Guillochon13} essentially confirm the numerical results of \citet{Lodato09} but extend their analysis to different penetration factors (\citealt{Lodato09} only considered the $\beta=1$ case). They also simulate partial disruptions, and give more solid estimates of the amount of mass loss and on the critical penetration for disruption, with respect to the earlier analysis of \citet{Khokhlov93} and \citet{Khokhlov93a}. The close agreement of the results of these two papers despite the different numerical technique used, and the convergence tests presented in them, allow to put solid constraints on the outcome of the disruption phase (which is anyway the simplest phase to simulate numerically).

More recently, \citet{Mainetti17} compare different codes to determine the critical penetration for total disruption. They use two different implementations of SPH, a moving mesh code and also compare their results to those obtained with FLASH by \citet{Guillochon13}. They find a broad agreement between the various codes and set $\beta_{\rm crit}=0.92$ and $2.01$ for the two cases $\gamma=5/3$ and $\gamma=4/3$, respectively. 

From the numerical point of view, a couple of papers have recently included the stellar magnetic field in MHD simulations. This has been done both with grid based codes by \citet{Guillochon17} and with SPH \citep{Bonnerot17b}. Interestingly, unlike earlier calculations of magnetic field amplification during tidal encounters, the results of these two papers are consistent with each other, probably due to the enhanced divergence cleaning method used in the SPH code used by \citet{Bonnerot17b}.

Finally, some simulations now go beyond simple polytropic models for the star and use instead MESA based stellar structures. One example is \citet{Golightly19b} who introduced MESA based density profiles in their SPH simulations, emphasising their role in the shape of the fallback rate. \citet{LawSmith19} introduce realistic stellar models in their FLASH simulations, which allows to track the compositional evolution of the debris stream.

For deeply penetrating events, the tidal compression can induce nuclear reactions in the disrupted star and many studies have studied this problem, as discussed above in Section 4. 

\subsection{Disc formation}

Following the stellar disruption, the debris evolves into an elongated stream that keeps orbiting the black hole with approximately half of it being bound and falling back towards the compact object while the rest gets unbound and escapes. When the bound gas returns to pericenter, it starts forming an accretion disc around the black hole. This process is initiated by a shock occurring when the tip of the stream intersects with the later-arriving material due to a modification of its trajectory by relativistic apsidal precession. The physics at play in the phases of stream evolution and subsequent disc formation are described in detail in the \flowchap{} in this book and we focus here only on the various numerical methods used to study them.

Numerical studies of the stream evolution have been carried out until the first debris falls back to pericenter using both SPH \citep{coughlin2015} and grid-based codes \citep{Guillochon14}. The former investigation found that the stream can fragment into self-gravitating clumps. While the perturbation at the origin of this collapse is likely numerical in this work, several processes exist that can sustain this fragmentation including periodic density variations inherited from the passage of the star at pericenter \citep{coughlin2016}. Such simulations do not require a large computational cost.

Following the process of disc formation from this stream when it comes back to pericenter is, however, very numerically challenging. Most importantly, an accurate numerical treatment of the passage of the stream at pericenter requires one to properly resolve its thin transverse profile. Reaching such a resolution for each part of the stream represents in itself a large computational overhead because of its very elongated profile. Also, because of such elongation, the evolutionary time at the head and tail of the stream differ substantially, so that while the time-step criterion is usually determined at pericenter, the system needs to be evolved over the long dynamical time-scale at apocenter. In SPH, such disparity of time-scales is  alleviated substantially by the employment of individual particle time-steps \citep{Bate95}. Remarkably, the issue of numerical resolution appears to be present independently of the numerical method used. Using SPH to explore this phase of evolution requires an unmanageably large number of particles. It has been tackled early on by \cite{Ayal00}, but it is now clear that their main results are flawed by a resolution much lower than would be required to accurately follow this process. Perhaps the most important artifact is that the majority (75\%) of the stellar mass gets unbound during the passage of the stream at pericenter. This effect is likely due to artificial dissipation taking place at this location due to the unresolved stream thickness. It is also unsurprising because they used a few thousand SPH particles while more recent (unpublished) investigations find that a thousand times more particles is still not enough to get rid of similar spurious numerical effects. Grid-based codes encounter the same type of numerical issues. While this numerical method could in principle study this process by using a more refined grid near pericenter, such a simulation has not yet been carried out, possibly due to the too large computational cost required. This issue may be resolved by the advent of new dedicated computational methods such as those discussed in the \fmodchap{} by Krolik et al in this book but this has so far not been the case.

One method used to gain insight into the disc formation process while avoiding the above computational burden was to study the self-crossing shock by means of local simulations. The passage of the stream near pericenter is followed analytically assuming that its orbit remains ballistic, which can be carried out by taking into account relativistic apsidal precession but also spin-related effects. This preliminary calculation allows to determine the properties of the two components of the stream when they reach the collision point, which is then used to initialize the local simulation of the self-crossing shock. This approach was used early on by \citet{kim1999}, who used an approximate treatment of radiation to evaluate the amount of thermal energy injected into the shocked gas that is able to diffuse out. More recently, this technique has been employed assuming gas adiabaticity \citep{lu2020} and making use of a more accurate algorithm for radiation transport \citep{Jiang2016}.

Global simulations of the disc formation process have been performed but all of them rely on important simplifications made in order to alleviate the computational burden discussed above that arises in the physically realistic case. Most of these simplifications aim at artificially reducing the size of the stream, either by decreasing the black hole mass to $M_{\rm bh} \approx 10^3 M_{\odot}$ or considering stars on elliptical rather than parabolic orbits with eccentricities in the range $0.8 \leq e \leq 0.97$. Such simulations have been carried out using both grid-based codes and the SPH technique and using various treatments of general relativistic effects from full GR to simpler gravitational potentials (as discussed in more details in Section \ref{sec:rel}). 
Early simulations, including those by \citet{Rosswog09}, have investigated the early phases of disc formation, although not in a detailed manner. The first one to follow the disc formation process in its entirety has been carried out by \citet{Hayasaki13} who consider an eccentric stellar trajectory with $e =0.8$ and $\beta=5$ and assumed an isothermal evolution for the gas to approximate efficient gas cooling. This work was improved by \citet{Bonnerot16} who focused on more eccentric stellar orbits with $e =0.95$ and a grazing penetration factor $\beta =1$. In addition, this study investigated the influence of inefficient gas cooling corresponding to an adiabatic evolution of the gas. These works treat relativistic apsidal precession through the use of pseudo-Newtonian potentials. The hydrodynamics of disc formation has also been investigated with a grid-based code that treats general-relativistic effects exactly by \citet{Shiokawa2015} considering a parabolic stellar trajectory around a black hole of mass $M_{\rm bh}=500 M_{\odot}$ and assuming adiabaticity. The results obtained are similar to the earlier ones concerning the timescale and final outcome of the disc formation process. More recently, \citet{bonnerot2019} have used another simplifying strategy that consists of modelling the self-crossing shock by an injection of outflowing gas inside their computational domain, which allows them to follow disc formation for astrophysically realistic parameters of the problem, most importantly a black hole mass $M_{\rm bh} = 2.5 \times 10^6 M_{\odot}$ and a parabolic stellar orbit. 

Two further improvements have been made more recently. The first shows that the influence of the black hole spin can delay the initial self-crossing shock through Lense-Thirring precession. This additional effect has been investigated by \citet{Hayasaki16} who consider bound stars and a treatment of spin-related effects through a gravitational potential with post-Newtonian corrections. The second advance is the inclusion of magnetic fields inside the stellar debris, which was carried out in the numerical work by \citet{Sadowski16}. These authors use a hybrid approach that consists of simulating the stellar disruption with SPH and the ensuing disc formation process using a grid-based code after adding magnetic fields to the debris. Most importantly, they were able to study the relative influence of magnetic fields and gas dynamics at producing the turbulence at the origin of angular momentum transport through the newly-formed disc.

Recently, \citet{Chan19} have simulated the interaction of the debris stream with a pre-existing accretion disc, when the disruption occurs in an AGN. They have used the grid-based code Athena++ and find that the stream can excite significant inflow in the disc (strongly super-Eddington), while the emergent radiation is expected to be close to the Eddington value.

\subsection{Accretion}

After a disc is formed out of stellar debris, viscous or magnetic processes in the disc can effectively transport the angular momentum of gas outwards \citep{Shakura73,Balbus91}, allowing it to accrete onto the black hole and produce emission. If the black hole mass is smaller than $\sim 10^8~M_\odot$ and the disc formation process is prompt, the accretion rate is expected to resemble the fallback rate, which exceeds the Eddington accretion rate for $\sim 1$yr around peak and then decays to below the Eddington level. We refer to the Chapter on Accretion Physics in Tidal Disruption Events by Dai et al in this book for the details of physics involved in this process and only focus on the development of numerical studies on TDE accretion here.    

While various hydrodynamical simulations have been employed to improve our understanding on the TDE disruption and disc formation processes as discussed in 5.1 and 5.2, there have been much fewer numerical studies on the TDE accretion process. One main reason for this is that we do not know the exact initial disc configuration to use, since currently we still cannot fully follow the long-term debris stream evolution until a disc is fully assembled without making simplifications. For example, in the simulations by \citet{Shiokawa2015} and \citet{Bonnerot16}, the assembled discs have different eccentricities and structures, due to the different initial set-up used to reduce the dynamical range. Also, while local simulations of debris stream-stream collision such as \citet{Jiang2016} show that such collisions can induce outflows and therefore the accretion rate could be lower than the fallback rate, an accurate measure of the lost debris fraction during the disc formation process can only be obtained using global simulations which include the black hole gravity.

The other reason for the slow progress is that numerical tools needed to study super-Eddington accretion, which is assumed to happen around the peak of most TDEs, have not been developed until a few years ago \citep[e.g.,][]{Ohsuga05, Sadowski14, Jiang14, McKinney14}. Unlike the accretion discs at low accretion rates, super-Eddington accretion discs, puffed up by large radiation pressure, are geometrically and optically thick. Therefore, the disc gas dynamics is affected by radiation, and the radiation emitted by disc inner region is heavily reprocessed in the outer disc and optically thick wind. While magnetohydrodynamical (MHD) codes are needed to study optically thin accretion discs, more advanced radiation-MHD (RMHD) codes are further needed to study optically thick super-Eddington discs. There are a few differences between these codes. Some codes are pseudo-Newtonian \citep{Ohsuga05, Jiang14} while others are general-relativistic \citep{Sadowski14, McKinney14}. Also the radiative transport scheme used in these codes are different (see section 4). However, the simulations of super-Eddington discs using these codes have produced qualitatively consistent results, showing that strong outflows can be produced and the escaped luminosity can greatly exceed the Eddington limit \citep[e.g.,][]{Jiang14, McKinney15}, which are very different from the predictions from analytical modelings \citep[e.g.,][]{Abramowicz88}.

Recently \citet{Dai18} perform the simulation of a circular, super-Eddington disc mimicking discs formed in TDE using the code \textsc{Harmrad} \citep{McKinney14}, assuming that such discs can circularize promptly and strong magnetic field threads the disc. The simulation results, post processed with another state-of-the-art Monte-Carlo radiative transfer code \textsc{Sedona}\citep{Kasen06, Roth16}, show that the anisotropic super-Eddington disc-wind structure can naturally lead to X-ray dominant or optical dominant emission from different viewing angles.  \citet{Curd19}, using the code \textsc{Koral} \citep{Sadowski14}, further study TDE super-Eddington discs around black holes with different spin parameters or magnetic flux strength. 

It is also worth noting that the current generation of GRRMHD codes are extremely computationally expensive. Therefore, it is inefficient to use such codes to follow the TDE disc evolution over the fallback time scale. Also, as the accretion rate approaches the Eddington accretion rate, the disc structure will become thin and the viscous timescale will grow longer, putting more challenges to simulating the transition from the super-Eddington regime to the thin disc regime. Here, one may instead just adopt the analytical solution of standard thin discs, although the classical question on whether or how such discs withstand thermal instability still remains to be solved by future numerical work.

\section{Conclusions and outlook}
\label{sec:conclusions}

Numerical simulations have played a major role over the years in sharpening our understanding of the dynamics and hydrodynamics of tidal disruption events. Here, we have presented the various methods that have been employed for this task, from Lagrangian particle based methods, such as SPH, to grid-based codes. We have discussed how such different methods have been applied to the various stages of the process of tidal disruption, and emphasised their successes and their limitations. Table \ref{tab:1} shows a ``map'' of the current effort in simulating TDEs in terms of different codes, treatment of thermodynamics and general relativity, etc. We hope this map can be useful to guide the reader into the field of numerical TDEs.

For what concerns the disruption phase, it is fair to say that current codes behave reasonably well. Numerical convergence is reached easily (already using $\sim 10^5$ particles, for SPH simulations) and it is encouraging to see that the results obtained with different methods agree well \citep{Lodato09,Guillochon13,Mainetti17}. Care should be taken in this context only for very deeply penetrating events, for which relativistic effects and thermonuclear detonation (for white dwarfs disruption) could play an important role. As a result, most of the numerical work done in this context, rather than improving the numerics, is aimed at including progressively more complex geometries, for example exploring disruption by binary black holes \citep{Coughlin17,Vigneron18} or the role of stellar spin \citep{Golightly19,Sacchi19}, or that of realistic stellar structure \citep{LawSmith19}.

\begin{table}
\caption{Map of numerical methods.  I: {\bf Method}, A: Affine, SPH, MM: Moving Mesh, G: Fixed Grid, AMR: Fixed grid w/ AMR, Log: Fixed Grid w/ Logarithmic Radius; II: {\bf Frame}, Loc: Local, Gl: Global; III: {\bf BH Gravity}, Newt: Newtonian, GR: General relativistic (either approximate or exact); IV: {\bf EOS}, Id: Ideal gas, MESA, H: Helmholtz, O: Other; V: {\bf MHD}, Y: Yes, N: No; VI: {\bf Radiation}, Y: Yes, N: No; VII: {\bf Nuclear reactions}, Y: Yes, N: No; VIII: {\bf Self-gravity}, Y: Yes, N: No; IX: {\bf Gravitational Waves}, Y: Yes, N: No; X: {\bf Trajectory}, P: Parabolic, E: Elliptical, Var: Variety, NR: not relevant}
\label{tab:1}       
 \begin{tabular}{p{3.75cm}cccccccccc}
\hline\noalign{\smallskip}
Paper & I & II & III & IV & V & VI & VII & VIII & IX & X\\
\noalign{\smallskip}\hline\noalign{\smallskip}
\citet{Carter82,Carter83,Carter85} &  A & Loc & Newt & Id & N & N & Y & Y & N & P\\
\citet{Luminet89a,Luminet89b} & A & Loc & GR & O & N & N & Y & Y & N & P\\
\citet{casalvieri2006} & A  & Loc & Newt & Id & N & N & N & N & Y & Var\\
\citet{Nolthenius82} & SPH & Gl & Newt & Id & N & N & N & Y & N &  P \\
\citet{Evans89} & SPH & Gl & Newt & Id & N & N & N & Y & N & P \\
\citet{Laguna93} & SPH  & Gl & GR & Id & N & N & N & Y & N & P\\
\citet{Ayal00} & SPH & Gl & GR & Id & N & N & N & Y & N & P\\
\citet{Lodato09} & SPH & Gl & Newt & Id & N & N & N & Y & N & P\\
\citet{Bonnerot16} & SPH & Gl & GR & Id & N & N & N & Y & N & E\\
\citet{Bonnerot17} & SPH & Gl & Newt & Id & Y & N & N & Y & N & P\\
\citet{coughlin2015} & SPH & Gl & Newt & Id & N & N & N & Y & N & P\\
\citet{Liptai19} & SPH & Gl & GR & Id & N & N & N & Y & N & E\\
\citet{Anninos2018} & MM & Gl & GR & O & N & N & Y & Y & Y & P\\
\citet{Goicovic19} & MM & Gl & Newt & Id/MESA & N & N & N & Y & N & P\\
\citet{Marck1996} & G & Loc & GR & Id & N & N & Y & Y & N & P \\
\citet{Khokhlov93,Khokhlov93a} & G & Loc & Newt & Id & N & N & N & Y & N & P\\
\citet{Frolov1994} & G & Loc & GR & Id & N & N & N & Y & N & P\\
\citet{Diener97} & G & Loc & GR & Id & N & N & N & Y & N & P\\
\citet{Brassart2008} & G & Loc & Newt & N & N & N & N & N & N & P\\
\citet{Cheng2013} & G & Loc & GR & Id & N & N & N & Y & Y & P\\
\citet{Cheng14} & G & Loc & GR & Id & N & N & N & Y & N & P\\
\citet{Guillochon2009} & AMR & Loc & Newt & Id & N & N & N & Y & Y & P\\
\citet{Guillochon13}  & AMR & Loc & Newt & Id & N & N & N & Y & N & P\\
\citet{Manukian2013}  & AMR & Loc & Newt & Id & N & N & N & Y & N & P\\
\citet{MacLeod2012}  & AMR & Loc & Newt & Id/MESA & N & N & N & Y & N & P\\
\citet{MacLeod2013}  & AMR & Loc & Newt & Id/MESA & N & N & N & Y & N & E\\
\citet{Guillochon2017} & AMR & Loc & Newt & Id & Y & N & N & Y & N & P\\
\citet{Law-Smith2017} & AMR & Loc & Newt & Id/MESA & N & N & N & Y & N & P\\
\citet{LawSmith19} & AMR & Loc & Newt & MESA/H & N & N & N & Y & N & P\\
\citet{Haas2012} & AMR & Gl & GR & Id & N & N & N & N & Y & P\\
\citet{East2014} & AMR & Gl & Newt & N & N & N & N & Y & Y & P\\
\citet{Shiokawa2015} & Log & Gl & GR & Id & N & N & N & N & N & P\\
\citet{Sadowski16} & Log & Gl & GR & Id & Y & N & N & N & N & E\\
\citet{bonnerot2019} & SPH & Gl & GR & Id & N & N & N & N & N & P \\
\citet{Jiang2016} & G & Loc & Newt & Id & N & Y & N & N & N & P \\
\citet{lu2020} & G & Loc & Newt & Id & N & N & N & N & N & P \\
\citet{Roth16} & G & Gl & Newt & O & N & Y & N & N & N & NR \\
\citet{Dai18} & Log & Gl & GR & Id & Y & Y & N & N & N & NR \\
\citet{Curd19} & Log & Gl & GR & Id & Y & Y & N & N & N & NR \\
\citet{Tanikawa2017,TanikawaMar2018,TanikawaMay2018} & SPH & Gl & Newt & H & N & N & Y & Y & N & P\\
\citet{Kawana2018} & SPH & Gl & Newt & H & N & N & Y & Y & N & P\\
\citet{Mainetti17} & SPH & Gl & Newt & Id & N & N & N & Y & N & P\\
\citet{Steinberg19} & MM & Loc & Newt & Id & N & N & N & Y & N & P\\
\citet{Yanilewich19} & MM & Gl & Newt & Id & N & N & N & Y & N & P\\
\citet{gafton2019} & SPH & Gl & GR & Id & N & N & N & Y & N & P\\
\citet{Tejeda17} & SPH & Gl & GR & Id & N & N & N & Y & N & P\\
\noalign{\smallskip}\hline
\end{tabular}
\end{table}


The subsequent evolution of the debris appears to be more problematic. For several years, it has been assumed that the debris would evolve according to simple Keplerian dynamics, with effects due to pressure and self-gravity being negligible. However, it has been recently shown that this may not be the case \citep{coughlin2015,coughlin2016} and the stream may develop (in the case of the disruption of giant stars) hydrodynamic instability \citep{Bonnerot16b} (although note that the instability criterion used by \citealt{Bonnerot16b} assumed an incompressible flow). The extent to which numerical effects can affect the evolution in this regime is still unclear: for example, \citet{coughlin2016} find, rather surprisingly, that increasing the resolution of their simulations (and thus improving the ability to resolve small scale fragmentation) actually reduces the tendency of the stream to fragment due to gravitational instability. 

 The formation of a disc or an accretion flow from the fallback material does present difficulties, especially for parabolic stellar orbits.  This is because the gas density at the beginning of the fallback is very low.  Application of adaptive methods (such as SPH) would result in very low spatial resolution, leading to unphysical behaviour.  In some sense, it is disappointing that no simulation to date has been able to follow the whole process of stellar disruption and disc formation for the standard case of a solar mass star on a parabolic orbit, being disrupted by a SMBH.  Additionally, this phase is expected to strongly depend on relativistic effects, such as apsidal and Lense-Thirring precession.  These effects still need to be properly accounted for in numerical schemes, although recent work has extended SPH to a general relativistic frame, by solving the equation of motion on a curved, static metric \citep{Liptai19}. 

Finally, the proper accretion phase is most naturally treated in the context of GR (radiation) MHD simulations. Simulations such as \citet{Dai18} have illustrated the basic properties of TDE discs in the super-Eddington phase. However, if one wants to fully understand the dynamical evolution of the disc as well as the reprocessed emission, there are a few remaining bottlenecks. 1) The current approach of treating the disc formation and accretion stages separately introduces an intrinsic uncertainty for the proper initialization of the disc when studying the TDE accretion phase. 
2) Close to the black hole, the disc structure is regulated by the accretion process. Typically the disc inflow equilibrium can only be established for a few tens of gravitational radii around the black hole in GRRMHD simulations which are computationally expensive. \citep[In] [this equilibrium can be established for hundreds of gravitational radii because the viscous timescale is shorter for a disc threaded by stronger magnetic flux.]{Dai18}  This problem will only become more severe when studying the disc transitioning from the super-Eddington regime to the sub-Eddington regime, when the disc structure becomes thinner and its viscous timescale grows longer. Therefore, the dynamical evolution of the disc over the whole fallback timescale can be hard to simulate. 3) The current generation of GRRMHD codes all employ frequency-averaged absorption opacity to calculate the gas-radiation coupling. In order to make this calculation more accurate and especially to obtain TDE spectra directly from disc simulations, frequency-dependent radiative transfer physics (such as line transitions) should be incorporated into the GRRMHD codes, which remains as a goal for future studies.

In this Chapter, we have concentrated on current efforts in simulating TDE, specifically from the algorithmic point of view, and certainly more work is required. We identify the most pressing area of further progress to be: (a) a proper inclusion of general relativistic effects, and in particular of the role of black hole spin; (b) a better treatment of dissipative effects, with the aim of resolving the dynamics that leads to disc formation; (c) a more detailed inclusion of radiative effects, that have been rarely included in simulations. Additionally, improved hardware, such as GPU computing, will definitely offer new resources for future work. A detailed discussion of such issues can be found in the dedicated Chapter within this book, by Krolik et al.

%
%


\begin{acknowledgements}
The authors thank the Yukawa Institute for Theoretical Physics at Kyoto University.
Discussions during the YITP workshop YITP-T-19-07 on International Molecule-type
Workshop "Tidal Disruption Events: General Relativistic Transients" were useful to
complete this work.  The work of RMC was funded by a Nicholas C.
Metropolis Postdoctoral Fellowship and the Advanced Simulation Computing Physics and Engineering Program under the auspices of Los Alamos National Laboratory, operated by Triad National Security, LLC, for the National Nuclear Security Administration of U.S. Department of Energy (Contract  No. 89233218CNA000001). The research of CB was funded by the Gordon and Betty Moore Foundation through Grant GBMF5076.  JLD  is  supported  by  the  GRF  grant  from the Hong Kong government under HKU 27305119.

\end{acknowledgements}

\bibliographystyle{aps-nameyear}      
\bibliography{simmethods,Lodato}                
\nocite{*}

%
%



\end{document}